# Nuclide Production in $^{197}$Au, $^{208}$Pb, and $^{nat}$U Irradiated with 0.8-1 GeV Protons: Comparison with other Experiments and with Theoretical Predictions

Yu.E. TITARENKO[*], V.F. BATYAEV, E.I. KARPIKHIN, V.M. ZHIVUN, A.B. KOLDOBSKY, R.D. MULAMBETOV, S.V. MULAMBETOVA, S.L. ZAITSEV, S.G MASHNIK[1], R.E. PRAEL[1], K.K. GUDIMA[2], M. BAZNAT[2]

*Institute for Theoretical and Experimental Physics, Moscow, Russia*
*[1] Los Alamos National Laboratory, Los Alamos, NM 87545, USA*
*[2] Institute of Applied Physics, Academy of Science of Moldova, Chisinau, Moldova*

*We used the ITEP proton synchrotron U10 to irradiate isotopically-enriched $^{208}$Pb and $^{nat}$U thin targets with 1.0 GeV protons and $^{197}$Au thin targets with 0.8 GeV protons. More than 400 cross sections of the nuclides produced were measured using the direct γ-spectrometry method with a high-resolution Ge detector. The measured γ-spectra were interactively processed by the GENIE2000 code. The γ-lines were identified, and the cross sections determined, by the ITEP-developed SIGMA code using the PCNUDAT database.*

*The measured cross sections are compared with similar data obtained at GSI for kinematically inverse reactions of 1 GeV/nucleon $^{208}$Pb, 1 GeV/nucleon $^{238}$U, and 0.8 GeV/nucleon $^{197}$Au interacting with a hydrogen target and with the ZSR data on $^{nat}$Pb and $^{197}$Au irradiated with 1 and 0.8 GeV protons, respectively. Our results are on average ~ 10 to 20% higher than the GSI data. The measured data are analyzed with the LANL codes CEM2k+GEM2 and LAQGSM+GEM2 and with the INCL intranuclear cascade code from Liege merged with the GSI evaporation/fission code ABLA. The obtained results may help to find possible sources of some observed discrepancies, thereby increasing the reliability of the data (both measured and simulated) required for different accelerator applications.*

## INTRODUCTION

Development of high-energy hadron-nucleus interaction models and codes requires they are validated against reliable experimental data. Among such data are cross sections for residual nuclide production measured at different laboratories, for instance, at the Gesellschaft fuer Schwerionenforschung (Darmstadt, GSI), Zentrum für Strahlenschutz und Radioökologie (Universitaet Hannover, ZSR), and the Institute for Theoretical and Experimental Physics (Moscow, ITEP).

The ITEP and ZSR measurements of the product cross sections are based on proton irradiation of experimental samples and subsequent γ-spectrometric analysis (called hereafter the "direct kinematics" method)[1-5]. The GSI approach is fundamentally different and involves accelerated ion irradiation of a liquid hydrogen target with a subsequent recording of the mass and charges of the nuclei produced from the projectile-nuclei (the "inverse kinematics" method)[6-8].

In view of the above, it is topical to compare the different experimental data sets among each other and with results by modern theoretical codes, such as CEM2k+GEM2, LAQGSM+GEM2, and INCL+ABLA that have been recently developed at Los Alamos and Liege.

## COMPARISON BETWEEN THE "DIRECT" AND "INVERSE" KINEMATICS DATA

The features and results of the direct and inverse kinematics methods are described in detail in[1-9]. The results of three inverse kinematics experiments at GSI[6-8] may be compared with the direct kinematics results, namely, the $^{197}$Au[7)] and $^{208}$Pb[6)] GSI measurements at 0.8 and 1.0 GeV/nucleon, may be compared with the ITEP and ZSR data[2-5)], respectively, and the $^{238}$U GSI measurements at 1 GeV/nucleon[8)] may be compared with the ITEP data[9)]. Comparison of results by these two different methods has to allow for the fact that the direct kinematics method provides mainly cumulative cross sections and just a small fraction of independent cross sections, whereas the inverse kinematics method determines only independent cross sections. Also, the inverse kinematics method does not separate metastable and ground states that can be measured by the direct kinematics method. These circumstances were taken into account in our previous works[2,3)] where our ITEP experimental $^{208}$Pb and $^{197}$Au results were compared with the GSI and ZSR data. In the present work, as an addition to[2,3)] we compare our data on 1.0 GeV proton-irradiated $^{nat}$U with the GSI inverse kinematics measurements and also perform a detailed comparison of all three experimental data sets with theoretical results by simulation codes CEM2k+GEM2, LAQGSM+GEM2, and INCL+ABLA. Table 1 shows results of comparing the direct and inverse kinematics experimental data for all three reactions. The mean cross section ratios $<\sigma_{expA,i}/\sigma_{expB,i}>$ and the mean squared deviation factors $<F>$ were chosen as quantitative criteria for this comparison. Fig. 1 shows another comparison of these experimental data sets, namely for the residual product mass distributions. Considered together, the mass distributions and the two quantitative criteria make it possible to judge with a higher confidence both about the level of systematic differences in the compared data (the cross section ratio) and about the differences proper (the mean squared deviation factor).

---

[*] Corresponding author, E-mail: Yury.Titarenko@itep.ru

**Table 1.** Results of quantitative comparison between the direct and inverse kinematics experimental data.

| Compari-son | $<\sigma_{expA,i}/\sigma_{expB,i}>$* | | | $<F>$* | | |
|---|---|---|---|---|---|---|
| | $^{197}$Au + p 0.8GeV | $^{208}$Pb + p 1.0GeV | $^{nat}$U + p 1.0GeV | $^{197}$Au + p 0.8GeV | $^{208}$Pb + p 1.0GeV | $^{nat}$U + p 1.0GeV |
| GSI/ITEP | 0.82 | 0.89 | 0.77 | 1.40 | 1.30 | 1.83 |
| GSI/ZSR | 0.83 | 0.81 | - | 1.99 | 1.50 | - |
| ITEP/ZSR | 1.16 | 0.93 | - | 1.30 | 1.24 | - |

\* $<\sigma_{expA,i}/\sigma_{expB,i}>$, $<F>$ were calculated as:
$<\sigma_{expA,i}/\sigma_{expB,i}>=10**<R_i>$, $<F>=10**sqrt(<R_i^2>)$, where $R_i=\log_{10}(\sigma_{expA,i}/\sigma_{expB,i})$

The following conclusions can be drawn from results presented in Table 1:

- We see a stable correlation between the values of the cross section ratios obtained by the direct kinematics method at ITEP and ZSR on the one hand, and by the fundamentally different inverse kinematics method at GSI, on the other hand. Namely, the cross section values obtained by the inverse kinematics method for all three reactions studied here are on average ~10-20% lower compared with those obtained by the gamma-spectrometry method. Interesting correlations can be also traced for values of the factor $<F>$: We see that, on average, our ITEP data are closer with the ZSR measurements and agree a little worse with the GSI data, while the biggest disagreement is observed between the ZSR and GSI measurements.

- Fig. 1 shows a significant difference in the experimental mass distributions for Au in the region A~110-140, that is intermediate between the fission and spallation modes. The mass yields for A=113 and 127 (ZSR) and 121 (ITEP) measured by the direct kinematics method are significantly higher (up to an order of magnitude) in comparison to the values obtained by inverse kinematics method (GSI). We note that there is no ground to assume a methodological error in the gamma-spectrometry measurements of these products from p+Au, as the yields of nuclides with the same masses from p+Pb measured by the same method agree well with the GSI data.

The observed discrepancies are not yet completely clear to us and should be overcome by further measurements.

## THEORETICAL RESULTS

In the present work, we have simulated all studied measured cross sections with the Liege intranuclear cascade model INCL[10] in conjunction with the GSI evaporation/fission model ABLA[11] (INCL+ABLA) and using the improved Cascade-Exciton Model (CEM) code CEM2k[12] and the Los Alamos version of the Quark-Gluon String Model code LAQGSM[13] both merged[14] with the Generalized-Evaporation Model code GEM2 of Furihata[15]. Table 2 shows results of a quantitative experiment-to-simulation comparison, using the same criteria we compared different experimental sets in Table 1. Fig. 1 shows a qualitative comparison of calculation results with the data, presenting experimental and simulated mass distributions for all three reactions studied here.

**Table 2**. Quantitative comparison the ITEP, ZSR, and GSI experimental data with results by the LAQGSM+GEM2, CEM2k+GEM2, and INCL+ABLA codes.

| Data set | $<\sigma_{calc,i}/\sigma_{exp,i}>$ | | | $<F>$ | | |
|---|---|---|---|---|---|---|
| | LAQGSM | CEM2k + GEM2 | INCL + ABLA | LAQGSM | CEM2k + GEM2 | INCL + ABLA |
| | $^{197}$Au + p 0.8GeV | | | | | |
| ITEP | 0.69 | 0.85 | 0.63 | 1.83 | 1.63 | 2.07 |
| ZSR | 0.49 | 0.73 | 0.45 | 2.53 | 2.45 | 2.94 |
| GSI | 0.75 | 0.99 | 0.79 | 1.96 | 2.17 | 1.89 |
| | $^{208}$Pb + p 1.0GeV | | | | | |
| ITEP | 0.66 | 0.87 | 0.63 | 1.86 | 1.90 | 2.07 |
| ZSR | 0.55 | 0.71 | 0.57 | 2.26 | 1.94 | 2.30 |
| GSI | 0.68 | 0.83 | 0.84 | 2.10 | 2.58 | 1.88 |
| | $^{nat}$U + p 1.0GeV | | | | | |
| ITEP | 0.58 | 0.64 | 0.63 | 2.45 | 2.50 | 2.04 |
| GSI | 0.89 | 0.99 | 0.84 | 2.11 | 2.67 | 1.90 |

From the quantitative and qualitative comparisons presented in Table 2 and Fig. 1 one can see that, on the average, all three codes provide a reasonable good agreement with all three experimental data sets and none of the tested codes shows a decisive advantage over other codes for the reactions studied here.

At the same time, partial comparison of results for a given reaction or for production of specific nuclei demonstrates different degrees of agreement of tested codes with experimental data from different sets:

- The LAQGSM+GEM2 and INCL+ABLA codes give very similar calculated-to-experimental mean cross section ratios when compared with data from each of three groups (the mean difference is around ~40%, see Table 2). At the same time, the CEM2k+GEM2 code gives, as a rule, much lower values of systematic deviations (~10-20% on the average) that are comparable with the respective values for the "experimental" comparisons. On the average, looking both at mean cross section ratios and mean square deviation factors, the CEM2k+GEM2 code seems to agree better with all experimental data sets, though the calculated cross sections are on the average below the experimental data in all cases;

- On the average, we see that the INCL+ABLA code agrees better with the GSI data and not so well with the ITEP and ZSR measurements, and just an opposite situation for CEM2k+GEM2 and LAQGSM+GEM2.

- The observed systematic different agreement of different codes with different data sets suggests that we should not limit ourselves to comparisons with experimental data from only one type of measurements when developing and benchmarking nuclear reaction models and codes.

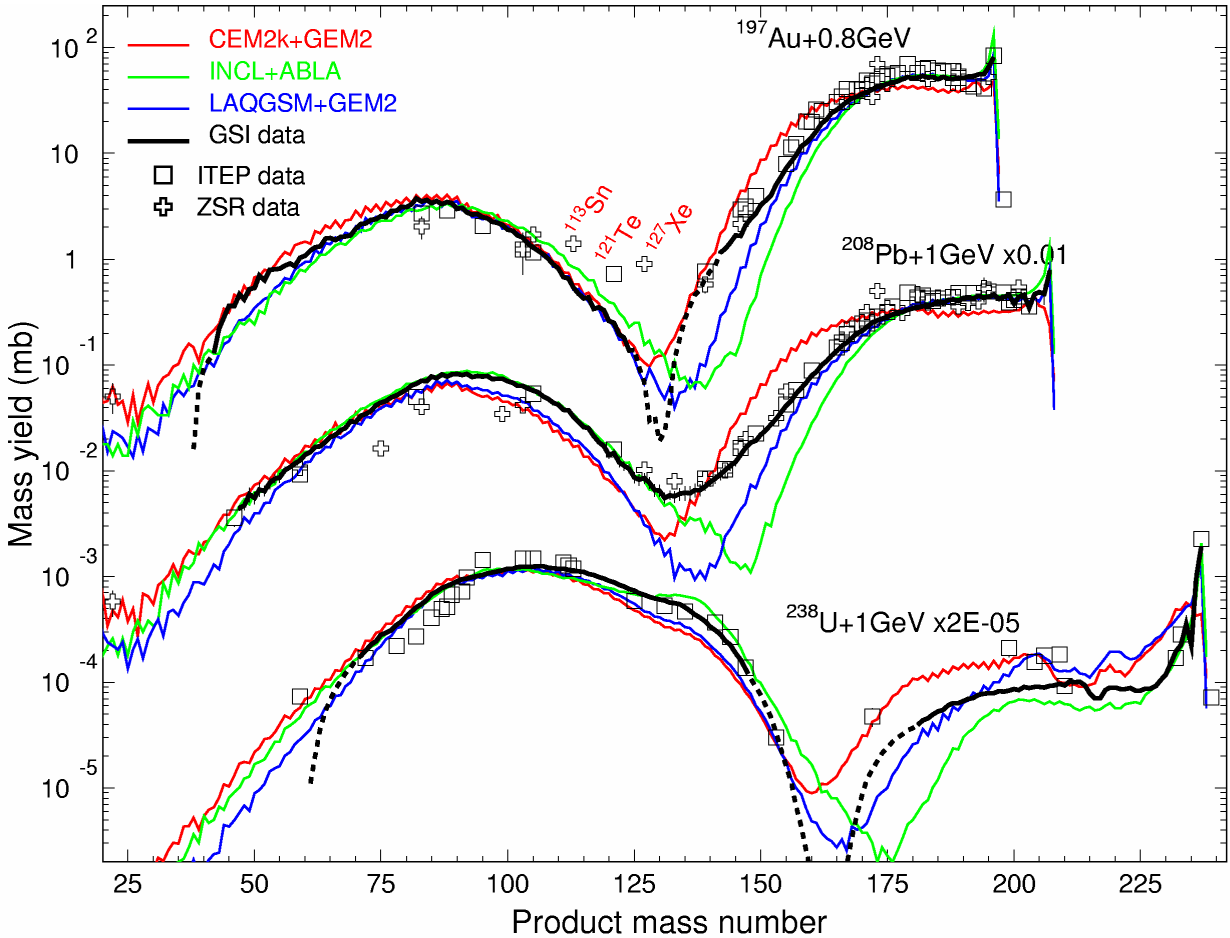

**Fig. 1**. Experimental and theoretical mass distributions of product nuclides from 0.8 GeV p + Au and 1.0 GeV p + Pb and U (The $^{208}$Pb GSI data are from [16]).

ACKNOWLEDGMENT

The authors are indebted to Drs. S. Leray and A. Boudard of CEA-Saclay for providing us with the latest version of the code INCL+ABLA and to Prof. K.-H. Schmidt and Dr. T. Enqvist of GSI-Darmstadt for providing us numerical data of the GSI measurements and fruitful discussions.

This work was partly carried out under the ISTC Projects # 839 and 2002 and was partly supported by the U.S. Department of Energy, Moldovan-U. S. Bilateral Grants Program, CRDF Project MP2-3045, and by the NASA Astrophysical Theory Program grant, Project #NRA-01-01-ATP-066.